\documentclass[prb,10pt]{revtex4-1}

\usepackage{amsmath}
\usepackage{amsfonts}
\usepackage{amssymb}
\usepackage{graphicx}
\usepackage{mathrsfs}
\usepackage{times}
\usepackage{color}
\usepackage{psfrag}
\usepackage{natbib}
\definecolor{violet}{rgb}{0.5,0,0.5}
\definecolor{blue2}{rgb}{0.5,0,1}
\usepackage{setspace}
\usepackage{graphicx,subfig}
\usepackage{array,calc}
\usepackage{braket}
\usepackage{bm}
\usepackage[justification=raggedright]{caption}

\newcommand{\di}{i}

\begin{document}


\title{Mechanics and dynamics of two-dimensional quasiperiodic composites}

\author{Danilo Beli$^1$} 
\author{Matheus I. N. Rosa$^2$} 
\author{Carlos De Marqui Jr.$^1$} 
\author{Massimo Ruzzene$^2$} 
\affiliation{$^1$S\~{a}o Carlos School of Engineering, University of S\~{a}o Paulo, Brazil}
\affiliation{$^2$Department of Mechanical Engineering, University of Colorado Boulder, USA} 




\date{\today}

\begin{abstract}
Periodic configurations have dominated the design of phononic and elastic-acoustic metamaterial structures for the past decades. Unlike periodic crystals, quasicrystals lack translational symmetry but are unrestricted in rotational symmetries, which leads to largely unexplored mechanical and dynamic properties. We investigate a family of continuous elastic quasicrystals with different rotational symmetry orders that are directly enforced through a design procedure in reciprocal space. Their mechanical properties are investigated as a function of symmetry order and filling fraction. Results indicate that higher order symmetries, such as 8-, 10- and 14-fold, allow for high equivalent stiffness characteristics that interpolate those of the constituent material while maintaining high levels of isotropy for all filling fractions. Thus, quasiperiodic composites exhibit more uniform strain energy distributions when compared to periodic hexagonal configurations. Similarly, nearly-isotropic wave propagation is observed over a broader range of frequencies. Spectral contents are also investigated by enforcing rotational symmetry constraints in a wedge-type unit cell, which allows for the estimation of band gaps that are confirmed in frequency response computations. The investigations presented herein open avenues for the general exploration of the properties of quasiperiodic media, with potentials for novel architectured material designs that expand the opportunities provided by periodic media.
\end{abstract}




\maketitle

\section{Introduction}

Quasiperiodic crytals, or shortly quasicrystals, are aperiodic structures with well-defined diffraction patterns exhibiting a set of sharp peaks, that are incompatible with translational symmetry, but still render long-range order ~\cite{LevineSteinhardt1984}. As such, quasiperiodic functions form a subset of deterministic aperiodic functions that have long been investigated mathematicians~\cite{Bruijn1981, Duneau1985, Gahler1986}. In physics, the interest in quasicrystals was perhaps ignited by the first observation of their natural occurrence in 1984 by Levine and Steinhardt~\cite{Shechtman1984}. In their seminal work, the authors provided evidence of a crystal with icosahedral point group symmetry, which is inconsistent with the translational symmetry of periodic crystals. Currently, it is known that quasicrystals may indeed exhibit rotational symmetry orders forbidden for periodic crystals, such as 5-, 7-, 8- and 10-fold symmetries~\cite{Lubensky1988, Steurer2008, Levi2011, Kraus2016}. The unique properties of quasicrystals has led to applications such as super-focusing \cite{Vardeny2013}, low-friction coatings \cite{Dubois1991}, thermal insulation \cite{Dubois2000}, and superconductivity \cite{Kamiya2018}, to name a few. In this work, we explore continuous elastic quasicrystals to understand their fundamental physical properties, and, furthermore, expand the range of possibilities for the design of architectured materials characterized by a broad range of symmetries and possibly superior performance.

In the context of architectured materials, a common goal is to design lightweight structures with high mechanical performance, that its, high stiffness with low mass density (or porosity)~\cite{Zheng2014, Rosario2017}. Isotropic stiffness is typically a desirable feature, in particular when the load directions are not known a priori~\cite{Latture2018}. Moreover, isotropy provides a relatively uniform distribution of strain energy, reducing stress concentration and the associated potential onsets of fracture ~\cite{Berger2017, Glacet2018}. Since regular materials generally lack such performance requirements at low densities, metamaterial designs have received attention in the recent research for such high performance materials~\cite{Zheng2014, Portela2020}. Although most studies focus on periodic metamaterials, the literature also includes quasiperiodic designs, for example in the form of truss lattices \cite{Chen2020, Wang2020} or foams \cite{Vidyasagar2018, Kumar2020}, which may be advantageous due to higher isotropy resulting from their higher order rotational symmetries~\cite{Chen2020}. However, the properties of continuous quasicrystals, such as the ones investigated in this paper, remain largely unexplored. Furthermore, while wave manipulation, filtering and attenuation capabilities of periodic metamaterials have been widely explored~\cite{Joannopoulos1997, Hussein2014, Ma2016, Cummer2016, Beli2018}, the study of the dynamic properties of quasiperiodic structures is still at an early stage. Quasiperiodic structures are notoriously difficult to handle due to the lack of translational symmetry, which precludes the application of Bloch's theorem for band structure estimations. Nonetheless, these structures may exhibit band gaps (or pseudo gaps)~\cite{Chan1998,Zoorob2000, Lai2002,King2007, Gei2010, Pal2019, Timorian2020}, in addition to other properties such as nearly isotropic wave propagation~\cite{Chen2020} and mode localization with~\cite{Jin1999, Bayindir2001} or without~\cite{Zhang2012, Jeon2017} defects. Quasicrystals are also known to exhibit remarkable spectral and topological properties, which are currently pursued in different fields such as condensed matter~\cite{prodan2015virtual,collins2017imaging,Duncan2020,chen2020higher}, photonics~\cite{kraus2012topological,vardeny2013optics}, acoustics~\cite{apigo2019observation,ni2019observation}, and mechanics~\cite{apigo2018topological,rosa2019edge,Pal2019,zhou2019topological,xia2020topological,riva2020adiabatic,rosa2020topological,riva2020edge,xia2020experimental,cheng2020demonstration}. 

Motivated by these contributions, we investigate the mechanical and dynamical properties of two-dimensional (2D) continuous elastic quasicrystals. The considered configurations consist of two-material composites, whereby the material distribution is defined by a design approach in the reciprocal space, or wavenumber domain~\cite{Lubensky1988, Widom2008}. The approach is employed to establish a family of configurations with desired rotational symmetry order, such as 4-, 6-, 8-,10- and 14-fold. We first examine their equivalent static properties, i.e. the variation of their equivalent stiffness as a function of symmetry order, volume fraction and direction. The results show that higher order symmetries that are forbidden for periodic crystals, present high equivalent stiffness (that interpolates between the stiffness of the two constituent materials) while ensuring high levels of isotropy for all volume fractions. The estimated properties also lead to nearly-isotropic wave propagation over a broader range of frequencies if compared with periodic counterparts, a behavior which is confirmed through numerical time-domain transient simulations. Finally, we conduct a dynamic analysis based on a wedge unit-cell with enforced rotational symmetry conditions, which provides estimations of band-gap frequencies as a function of volume fraction. Existence of such band-gaps is confirmed by forced response computations for finite domains.

The paper is organized as follows: following this introduction, Section~\ref{secdesign} outlines the strategy employed for the design of rotationally symmetric continuous quasicrystals. The numerical results for mechanical and dynamical properties are presented in Section~\ref{sec:Mechanical} and Section~\ref{sec:Dynamical}, respectively. Finally, Section~\ref{sec:Conclusions} summarizes the main findings of this work and outlines possible future research directions.


\section{Geometry description and analysis methods} \label{secdesign}

\subsection{Design of quasiperiodic composites in wavenumber space}

\begin{figure}
	\centering
	{\includegraphics[scale=0.600]{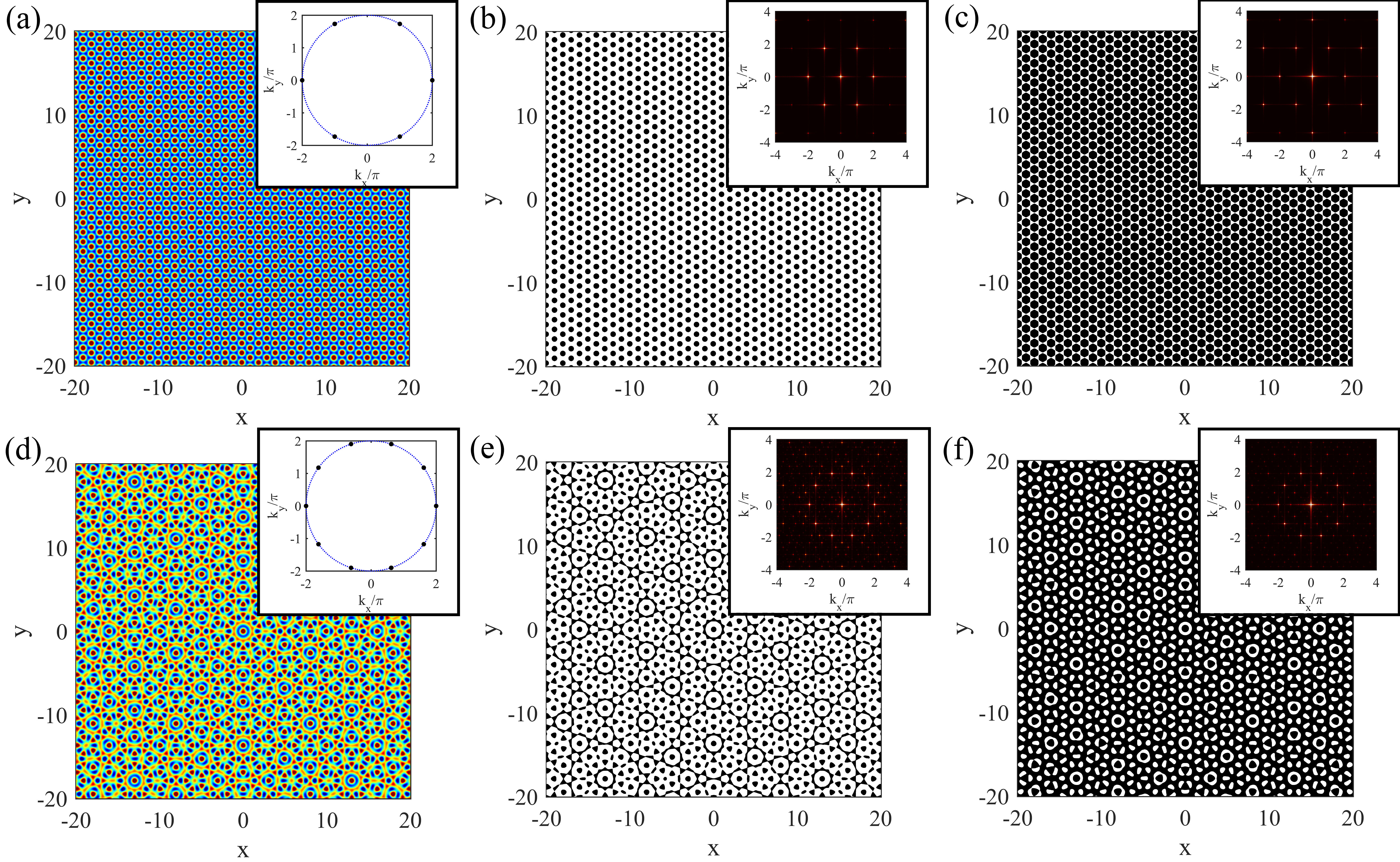}} 
	\caption{\label{fig:figure1} Design of quasiperiodic composites based on assignment of Bragg peaks in reciprocal space, exemplified by the 6-fold (hexagonal crystal, a-c) and by the 10-fold (dodecahedron quasicrystal, d-f) rotations. The real field (a, d) results from directly enforcing the Bragg peaks (black dots populating the circles in the insets). Examples of threshold distributions for ${\tt vf} = 0.30$ (b, e) and ${\tt vf} = 0.70$ (c, f), where white color represents the phase $\tt 0$ and the black color represents the phase $\tt 1$. Fourier transform of the threshold fields are depicted in the insets.} 
\end{figure}

Two-dimensional periodic and quasiperiodic composites are designed by means of a rational strategy that assigns specified characteristics of their reciprocal space ${\bf k} = [k_x, \,\, k_y ]\in \mathcal{R}^2$~\cite{Lubensky1988, Widom2008}. According to this strategy, the distribution of two materials is described by a function $\phi(\bf r)$, with ${\bf r} = [x, \,\, y ]\in \mathcal{R}^2$, where $x,y$ are cartesian coordinates in physical space. Such distribution is defined by directly assigning $N$ Bragg peaks in reciprocal space as pure points in the Fourier spectra. This generates $N$ peaks on a circle of assigned radius $k_R$ and separated by the angle $\theta=2\pi/N$ (Fig. \ref{fig:figure1}), where $N$ denotes the desired rotational symmetry of the composite. Thus, the function $\phi(\bf r)$ is given by
\begin{equation}
\phi({\bf r}) = \Re \left\lbrace \sum_{n = 0}^{N-1} \delta({\bf k}-{\bf k}_n) e^{\di {\bf k}_n \cdot {\bf r}}\right\rbrace = \sum_{n = 0}^{N-1} \cos({\bf k}_n \cdot {\bf r}) 
\label{eq:eq2}
\end{equation}
where $\delta$ is the delta function, which locates the wavenumber $\bf k_n$ of each peak, which is
\begin{equation}
{\bf k}_n = k_R [\cos \left( 2\pi n/N \right), \,\,  \sin \left( 2\pi n/N \right)], \quad  \quad n = 0,..., N-1.
\label{eq:eq1}
\end{equation}

This approach produces periodic crystals (i.e., structures with crystallographic point group symmetry), including 1D bilayer ($N=2$), square ($N=4$) and hexagonal ($N=6$) patterns, as well as quasicrystals with $N$-fold rotational symmetry orders, as illustrated in Fig. \ref{fig:figure1}(a, d). We remark that such design strategy always leads to even-fold symmetries in physical space, even when an odd number of Bragg peaks is chosen, since only the real part of the resulting field can be used to design the physical space. For instance, if an odd number of peaks is assigned (e.g., $N=5$), the real part of the resulting complex-valued function results in a real field of $2N-$fold symmetry order ($2N=10$), with $N$ extra peaks automatically assigned diametrically opposite to the initial $N$ peaks. 

A binary physical material distribution is obtained by applying a threshold to the continuum field $\phi(\bf{r})$ from Eq. \eqref{eq:eq2} by comparing it to a chosen level $\bar{\psi}$. This leads to a threshold distribution $\bar{\phi}({\bf r})$, where properties can assume only two values defined as phase $\tt 0$ (if $\phi({\bf r}) < \bar{\psi}$) or phase $\tt 1$ (if $\phi({\bf r}) > \bar{\psi}$), with a volume fraction defined by the ratio of volume of the space occupied by phase $\tt 0$ to the total volume, i.e.  ${\tt vf} = v_{\tt 0}/(v_{\tt 0}+v_{\tt 1})$. A few examples are presented in Fig. \ref{fig:figure1} for the 6-fold and 10-fold rotations with ${\tt vf} = 0.30$ and ${\tt vf} = 0.70$ (check \ref{sec:Extra1} for more examples). The Bragg diffraction of the threshold distributions $\bar{\phi}({\bf r})$ are also computed by using a spatial Fourier transform, which allows the comparison with the assigned Bragg peaks. Higher-order peaks (with small energy compared to design peaks) appear inside and outside the designed circle following the fold symmetry (see insets in Fig. \ref{fig:figure1}), which are related to the threshold procedure and to the trunctation of the physical space into a finite square domain. A central Bragg peak appear in all distributions, which is related to the average distribution produced by the threshold step.
However, even with the additional details, the Bragg diffraction of the threshold distributions maintains the $N-$fold symmetry order chosen by design. The threshold distributions are used next to design periodic and quasiperiodic materials with two constituent phases. To this end, mass density and Young's moduli distributions are defined as:
\begin{equation}
\rho({\bf r}) = \rho_{\tt 0} + \bar{\phi}({\bf r}) (\rho_{\tt 1}-\rho_{\tt 0}), \,\, E({\bf r} ) = E_{\tt 0} + \bar{\phi}({\bf r}) (E_{\tt 1}-E_{\tt 0})
\end{equation} 

\subsection{Analysis methodology}

We consider a 2D elastic domain in plane strain conditions, comprised of two constituent materials of high contrasting properties, i.e. $\rho_{\tt 1} = 10 \rho_{\tt 0}$ and elastic modulus $E_{\tt 1} = 10 E_{\tt 0}$. Unitary out-of-plane thickness and equal Poisson's ratio $\nu=0.33$ for the two phases are considered for simplicity.  Assuming linear elastic behavior, the domains are discretized within the {\tt COMSOL Multiphysics} \textsuperscript{\textregistered} environment.
The models where formulated using linear 2D shell elements with internal interpolation features, and considering meshes comprising 10 elements per wavelength $a$, where $a=2\pi/k_R$ is the fundamental wavelength corresponding to the circle of radius $k_R$ wherein the Bragg peaks of the material distributions are assigned. This discretization is found to properly describe the relevant mechanical and dynamical behavior considered in this paper. For the dynamic results, a normalized frequency $\Omega = \omega a/ c_L$ is employed throughout, where $c_L = \sqrt{{E_{avg}}/{\rho_{avg}}}$ is the longitudinal wave speed corresponding to averaged properties, i.e. $\rho_{avg} = (\rho_{\tt 0}+\rho_{\tt 1})/2$ and $E_{avg} = 2E_{\tt 0}E_{\tt 1}/(E_{\tt 0}+E_{\tt 1})$.


\section{Effective mechanical properties} \label{sec:Mechanical}

The effective mechanical properties of the quasiperiodic composites are investigated by using the standard mechanics approach ($\tt SMA$) \cite{Hollister1992} also known as averaging theory \cite{Nguyen2012}, whereby a sufficiently large domain is considered as a representative volume element ($\tt RVE$). The $\tt SMA$ is a well established method, where independent uniform longitudinal and shear displacements (or tractions) are applied at the $\tt RVE$ boundaries \cite{Portela2020, Nguyen2012}, as shown in \ref{sec:ContMechWave}. The resulting average stress (or strains) for these independent loads are used to derive the effective elastic tensor $\bar{\mathbf{C}}$ through the relationship $\bar{\sigma}_{ij} = \bar{c}_{ijkl} \bar{\varepsilon}_{kl}$, where $\bar{\sigma}$ is the resulted average stress tensor and $\bar{\varepsilon}$ is the resulted average strain tensor. Through the tensor transformation by means of direction cosines, the directional dependence of its coefficients are computed $\bar{c}'_{ijkl}(\theta) = \bar{c}_{mnpq} a_{im}a_{jn}a_{kp}a_{lq}$ \cite{Rosario2017, Portela2020, Hollister1992}, and hence, the directional effective elastic properties $\bar{E}(\theta)$, $\bar{G}(\theta)$ and $\bar{\nu}(\theta)$ are obtained from the elastic tensor according to the procedure described in \cite{Hollister1992, Nguyen2012} and as summarized in \ref{sec:ContMechWave}. In order to facilitate the effective properties comparison through chart plots, their values are projected in a basis aligned to the Cartesian coordinate system $\left\lbrace \vec{x},\vec{y} \right\rbrace$, where $\bar{E}_{x} = \bar{E}|_{\theta = 0}$,  $\bar{G}_{x} = \bar{G}|_{\theta = 0}$ and $\bar{\nu}_{x} = \bar{\nu}|_{\theta = 0}$. Based on the estimated effective properties, the anisotropic index measured by the Zenner ratio~\cite{Berger2017}:

\begin{equation}
\bar{\epsilon}_{\tt Z} = \bar{E}_{x}/[2 \bar{G}_{x}(1+ \bar{\nu}_{x})]
\end{equation}
is computed and employed to further characterize the considered class of bi-material composites. 

\begin{figure}[h!]
	\centering
	{\includegraphics[width=\textwidth]{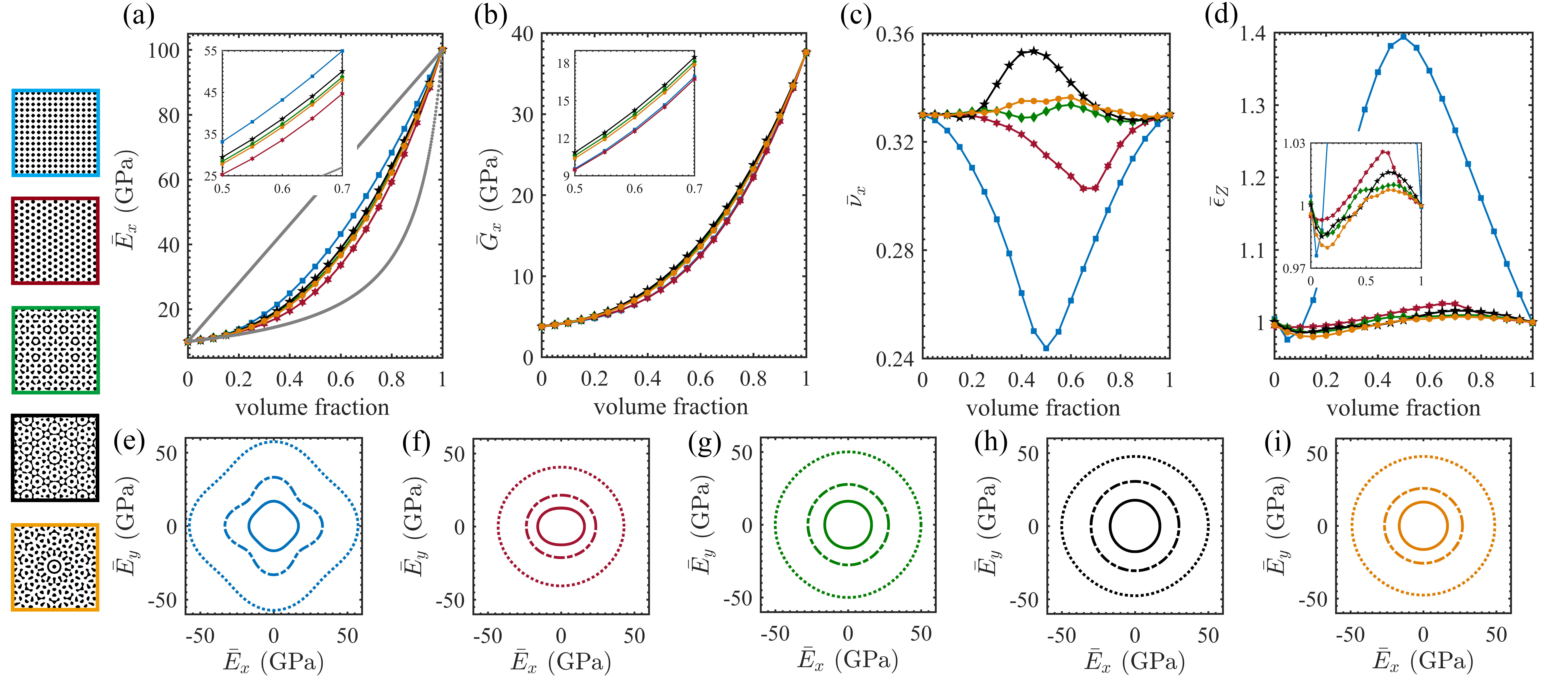}} 
	\caption{\label{fig:figure2} Chart plots showing the bounds of elastic properties in function of the volume fraction for 4- (blue), 6- (red), 8- (green), 10- (black) and 14- (yellow) fold rotations: elastic modulus with the Voigt and Reuss bounds in gray (a), shear modulus (b), Poisson coefficient (c) and Zener anisotropic ratio (d). Zoom of the elastic properties bounds at volume fractions between $0.50$ and $0.70$ to better distinguish the structural performance (a-b). The insets on the left column represent the fold rotations and not the real structure. Directional elastic contour for each fold symmetry (e-i) considering ${\tt vf} = \left[0.30 \ \ 0.50 \ \ 0.70 \right]$ (line, traces and points, respectively). The quasicrystal with 10-fold rotations presents the best structural performance with high total stiffness for the same volume fraction while the quasicrystals with 8- and 14-fold rotations have the nearest isotropic behavior for all volume fractions.} 
\end{figure} 

Results for effective mechanical properties ($\bar{E}_{x}$, $\bar{G}_{x}$, $\bar{\nu}_{x}$ and $\bar{\epsilon}_{\tt Z}$) for various periodic (4- and 6-fold) and quasiperiodic (8-, 10- and 14-fold) composites are shown in Figure~\ref{fig:figure2}(a-d). Variations of the effective properties with respect to the composites' volume fraction are presented in the form of chart plots~\cite{Zheng2014}, which also report the Voigt and Reuss bounds defined by the rule of mixtures \cite{Hill1952}. At low ($ {\tt vf} \leq 0.20$) or high (${\tt vf} \geq 0.80$) volume fractions, the properties are very similar for all fold symmetries, since they are dominated by a single phase. More significant differences, in particular with respect to the 4-fold crystal, are observed for the Poisson's ratio and Zener anisotropic ratio (Figure~\ref{fig:figure2}(c,d)). However, each symmetry order exhibits distinctive mechanical behavior for intermediate volume fractions (i.e, $0.20 < {\tt vf} < 0.80$). The square and hexagonal crystals are respectively characterized by the highest and the smallest elastic modulus $\bar{E}_{x}$ as a function of volume fraction, while the quasiperiodic composites bounds lie between them (Fig. \ref{fig:figure2}(a)). However, the quasiperiodic composites generally exhibit higher shear moduli, with the highest value observed for the 10-fold case (Fig. \ref{fig:figure2}(b)). The Poisson's ratio appears to have an opposite behavior for the periodic and quasiperiodic configurations, as it decreases with volume fraction for the former, and increases for the latter (Fig. \ref{fig:figure2}(c)). Also, the 4-fold configuration has the highest Poisson coefficient variation while the 8- and 14-fold composites maintain an almost constant value. Furthermore, the directionality plots in Fig. \ref{fig:figure2}(e-i) highlight that the quasicrystals provide an almost isotropic transition between the stiffnesses of the constituent phases, while the crystals (4-fold and 6-fold) show a stronger directional behavior, with the 4-fold case presenting the highest degree of anisotropy. These observations are also aligned with the Zener ratio reported in Fig. \ref{fig:figure2}(d). While the three quasiperiodic designs exhibit a similar almost-isotropic transition, we note that the 10-fold case is characterized by the highest elastic and shear modulus for each volume fraction, followed by the 8-fold configuration, and finally by the 14-fold case. 

\begin{figure}[h!]
	\centering
	{\includegraphics[width=\textwidth]{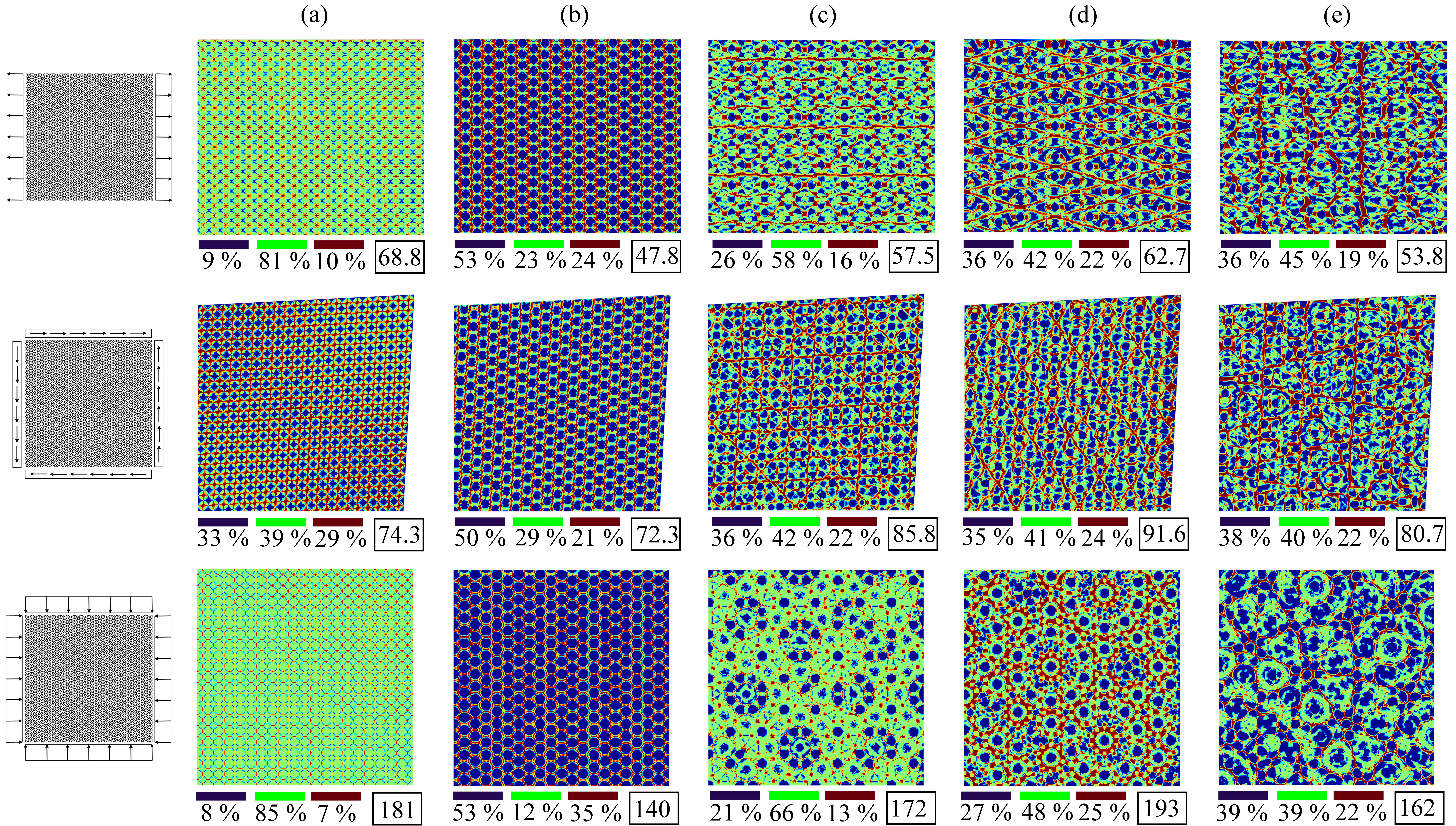}}  
	\caption{\label{fig:figure3} Strain energy distributions under uniaxial (first line), shear (second line) and hydrostatic (third line) loading for 4- (a), 6- (b), 8- (c), 10- (d) and 14- (e) fold rotations. The imposed displacements are small, but exaggerated to reveal their nature. The colors correspondent to zones with low (blue), medium (green) and high (red)  strain energy density in J/m$^3$, the correspondent percentage of the domain with these strain energy density levels are depicted below each case. The value inserted in the black rectangle corresponds to the total stored strain energy in J. All the results were computed for ${\tt vf} = 0.50$ and only a quarter of the domain (first quadrant) is shown due to symmetric nature of the patterns.} 
\end{figure}

Another important factor characterizing mechanical performance is the total strain energy (SE) given by a stress-strain integral within the domain, i.e. $\text{w}_{\Omega} = (1/{\Omega}) \int_{\Omega} (\sigma : \varepsilon) d \Omega$. For example, in energy-related applications, a configuration that stores more energy for the same volume fraction is considered a better harvester with promising applications in artificial muscles and soft robotics \cite{Zhan2020}. Overall, under the same loading conditions, higher values of elastic properties lead to higher strain energy storage. In addition to the absolute value of the total SE, its distribution in the composite (the strain energy density) is also relevant since it reveals regions which are not contributing/performing as well as parts with high density levels, and therefore more susceptible to failure. The strain energy of the composites is evaluated under uniaxial, shear and hydrostatic strain loads, which are independently applied at the boundaries of the $\tt RVE$ \cite{Rosario2017, Berger2017}. The resulting strain energy density patterns are displayed in Fig. \ref{fig:figure3}. Each distribution is normalized by the mean value of its strain energy density ($\text{w}_{m}$), and its visualization is improved by ranking the local strain energy (SE) density ($\text{w}_{l}$) in three levels: low  if $\text{w}_l < \text{w}_{m} - \text{w}_{s}/2 $ (in blue), medium if $\text{w}_{m} - \text{w}_{s}/2 \leq \text{w}_l \leq \text{w}_{m} + \text{w}_{s}/2 $ (in green) and high if $\text{w}_l > \text{w}_{m} + \text{w}_{s}/2 $ (in red), where $\text{w}_{s}$ is the standard deviation. Regarding the total SE levels (the numbers insert in the black rectangles), under uniaxial and hydrostatic loads, the 4-fold and 10-fold composites present the highest levels of storage, the 8-fold and 14-fold present medium levels of storage, while the 6-fold present the lowest. For shear loading, the 10-fold exhibits the highest level of storage, the 8-fold and 14-fold present medium levels, while the crystals (4-fold and 6-fold) present the lowest levels of storage. The smallest levels of storage for all the loads is observed in the 6-fold. These results are in agreement with the effective properties values showed in Fig. \ref{fig:figure2}, since the elastic coefficients are direct related to the total SE. In regards to the strain energy distribution (color maps), the most uniform pattern under axial and hydrostatic loadings is observed for the 4-fold crystal, where more than $80 \%$ of the domain presents a medium level of SE. For all the loads, the 6-fold crystal has more than $50 \%$ of the domain with low SE level, which indicates undesirable mechanical performance, since that sub-utilizes the material leading to high SE concentration in a few regions. The distribution performance of the quasiperiodic composites is similar under shear load; however, under axial and hydrostatic loads, the distribution performance is more uniform in the 8-fold case, followed by the 10-fold and then by the 14-fold.


\section{Dynamic behavior}\label{sec:Dynamical}

\subsection{Wave propagation and isotropy} \label{sec:Dynamical low}

\begin{figure}[h!]
	\centering
	{\includegraphics[scale=0.450]{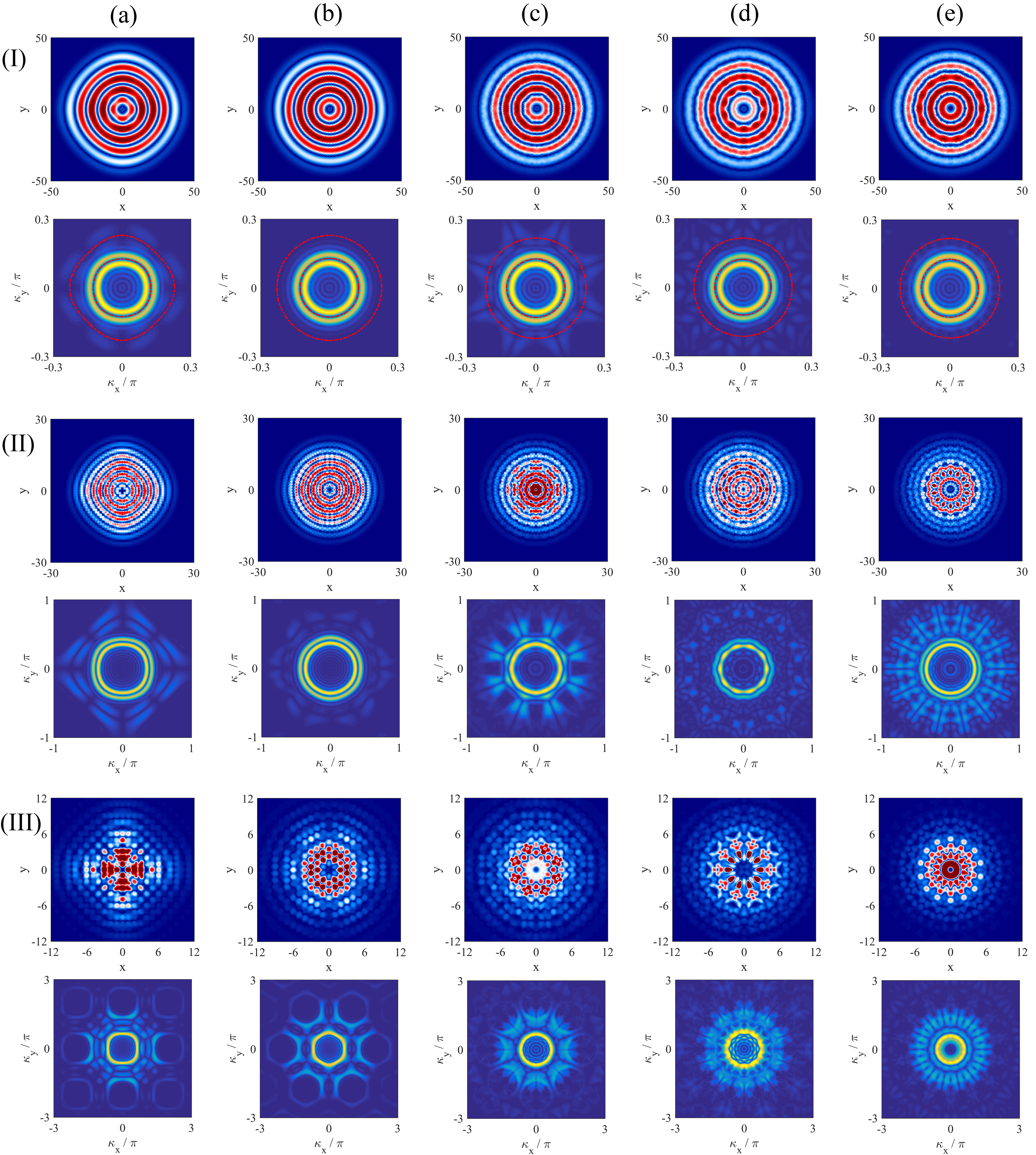}} 
	\caption{\label{fig:figure4} Snapshots of the magnitude of displacement field (i.e., $|{\bf u}(x,y,t)|$) and the correspondent wavenumber domain representation $|\hat{\bf U}(\kappa_x,\kappa_y,\omega)|$ for 4- (a), 6- (b), 8- (c), 10- (d) and 14- (e) fold rotations at $\Omega_{\tt I} \approx 0.08$ (I), $\Omega_{\tt II} \approx 0.24$ (II) and $\Omega_{\tt III} \approx 0.40$ (III). For low frequencies (case I), the wavenumber results are compared with theoretical iso-frequency dispersion contours (dashed red lines) obtained from the effective properties, where a good agreement between the results is observed. For increasing frequencies the 4- and 6-fold crystals present higher directional behavior, while the quasiperiodic crystals exhibit a higher degree of isotropy.  All the results were computed for ${\tt vf} = 0.30$.}  
\end{figure}

The wave propagation characteristics of the different composites are investigated through transient time domain simulations. The static behavior observed in the previous section is expected to shape the wave propagation of the composites at low frequencies (i.e., isotropic wave propagation). By increasing the excitation frequency, the periodic crystals (4-fold and 6-fold) exhibit higher directionality (i.e, wave motion occurs along preferential directions), while the quasiperiodic composites in general retain a higher degree of isotropy in their wave behavior. The simulations were performed using an omnidirectional excitation at the center of the domain, consisting of a small circle of radius $r=0.2a$ where radial forces were applied according to a band limited 5-cycle sine-burst excitation. The resulting displacement field ${\bf u}(x,y,t) = u(x,y,t)\vec{x} + v(x,y,t) \vec{y}$ is also represented in reciprocal space through a 3D Fourier transform  $\hat{\bf U}(\kappa_x,\kappa_y,\omega) = \hat{U}(\kappa_x,\kappa_y,\omega)\vec{\kappa}_x + \hat{V}(\kappa_x,\kappa_y,\omega) \vec{\kappa}_y$. At low frequencies, the longitudinal (P) and shear (S) branches are non-dispersive, and the components of the reduced elastic tensor can be related to the longitudinal and shear wave velocities (see details in \ref{sec:ContMechWave}). With the the wave velocities $\bar{c}_L$ and $\bar{c}_S$, the iso-frequency contours in reciprocal space for longitudinal waves $\bar{\kappa}_L = \omega / \bar{c}_L$ and for shear waves $\bar{\kappa}_S = \omega / \bar{c}_S$ can be estimated. This procedure is used to further validate the mechanical properties estimated in the previous section through the standard mechanics approach.

Simulation results for for all configurations and three different excitation frequencies $\Omega_I \approx 0.08$, $\Omega_{II} \approx 0.24$ and $\Omega_{III} \approx 0.40$ are summarized in Fig. \ref{fig:figure4}. These frequencies are selected based on a Bloch analysis conducted on the unit cell of the 4-fold lattice (see details in the Appendix C), and chosen to sample the longitudinal and shear wave branches before the first Bragg gap. For each excitation frequency, snapshots of the magnitude of displacement field $|{\bf u}(x,y,t)|$ at the same time instant are presented for each symmetry order (4-,6-,8-,10- and 14-fold in columns a to e, respectively), for a fixed volume fraction of ${\tt vf} = 0.30$. Also, the magnitude of the wavefield in reciprocal space at the excitation frequency $|\hat{\bf U}(\kappa_x,\kappa_y,\omega)|$ is displayed below each snapshot. For low frequencies (case I), wave directionality is observed primarily in the 4-fold crystal, where waves propagate preferentially along horizontal and vertical directions. The other composites display an almost circular wave front (i.e., wave motion without preferential direction), which is evidence of isotropic behavior. The contours for longitudinal waves and shear waves, computed using the effective properties, are displayed as dashed red lines and demonstrate a good agreement with the wavefield in reciprocal space. We note that longitudinal waves are dominant, with the shear waves presenting a lower amplitude, as expected from the type of excitation employed (radial forces in a circle). These results validate the equivalent properties obtained in the previous section, also suggesting that such approach may be useful in predicting the behavior of low-frequency waves in quasicrystals, where Bloch conditions cannot be applied. For higher frequencies (case II and III), the waves exhibit more directionality overall. In this regime, there is still a lack of understanding regarding the nature of wave propagation in quasiperiodic media since neither Bloch analysis or effective static properties can be used. Nonetheless, our numerical results indicate that, overall, the quasiperiodic structures retain a higher degree of isotropy when compared to the 4- and 6-fold crystals for increasing frequencies, with the 14-fold case exhibiting the highest degree of isotropy.


\subsection{Estimation of band gaps and frequency response} \label{sec:Dynamical high}

\begin{figure}[t!]
	\centering
	{\includegraphics[scale=0.550]{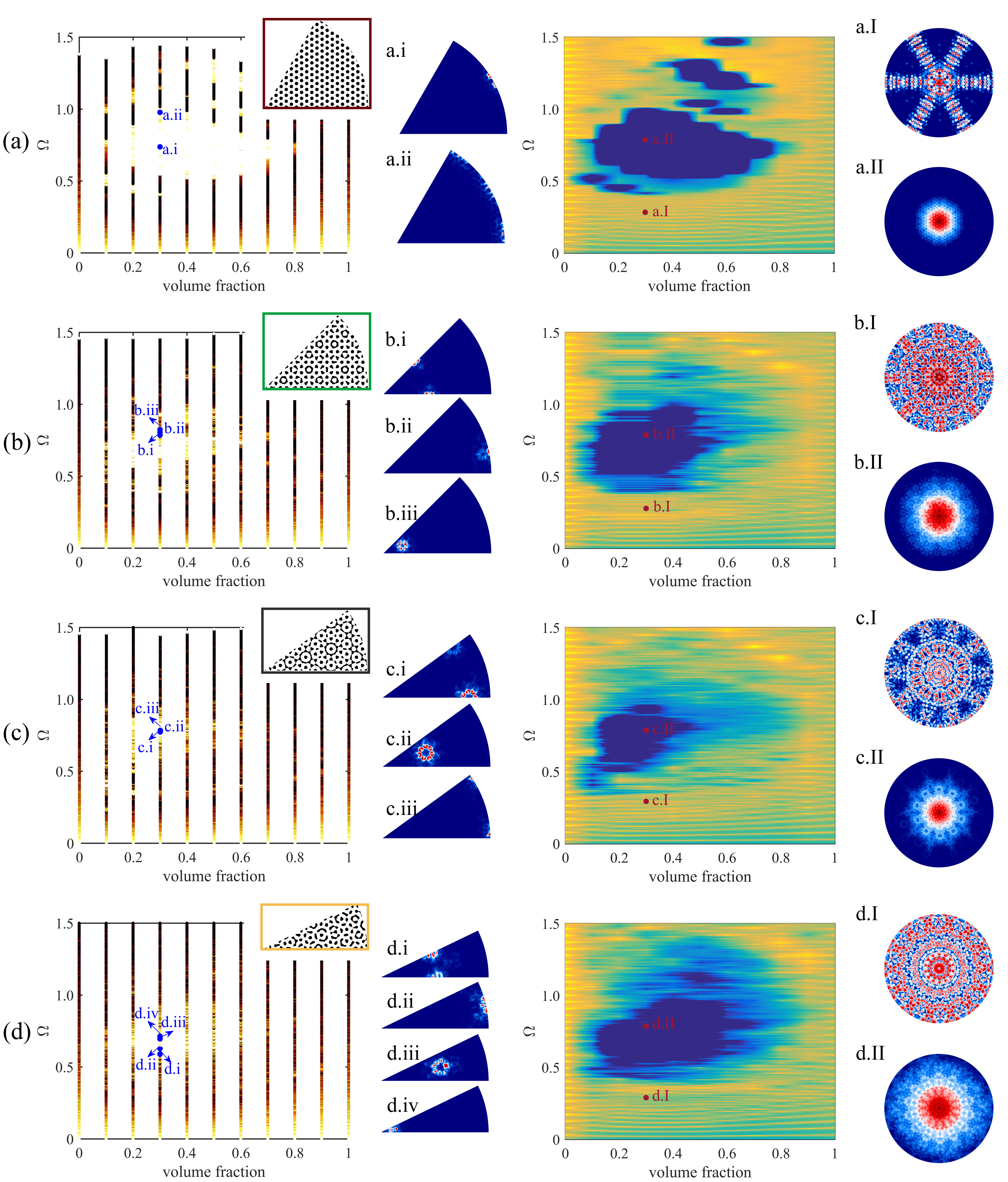}}
	\caption{\label{fig:figure5} Frequency spectrum of the circular sector (in the insets) corresponding to the 6- (a), 8- (b), 10- (c) and 14- (d) fold rotations symmetry (first column), where all modal periodicities $n$ are collapsed for each volume fraction. The density of states are depicted by the colors, where dark colors represent dense bulk bands and light colors signal low-density defect modes. The displacement field of the wedges (second column) showing some localized modes within the bulk band gap for ${\tt vf} = 0.30$ and $n = 2$. Maps of average harmonic displacement response of the complete structure by changing the volume fraction (third column) with the displacement fields (fourth column) before the band gap (I) and showing the wave attenuation inside a band gap zone (II) for ${\tt vf} = 0.30$. A great agreement between the frequency spectrum and the forced response is achieved.} 
\end{figure}

The spectral properties of the quasiperiodic composites are investigated next with the goal of evaluating the existence of band gaps. As previously mentioned, Bloch's theorem cannot be directly applied in quasiperiodic media due to the lack of translational symmetry. Although some works rely on periodic approximants~\cite{apigo2018topological,rosa2019edge,Pal2019,xia2020topological} to overcome such limitation, to the best of our knowledge, these methods are not application for 2D domains investigated here. To simplify computations, we consider wedge-shaped unit cells consisting on the smallest circular sectors defined by the $N-$fold rotational symmetry (i.e., $\theta_{sec} = 2\pi/N$). Cyclic symmetric periodic conditions ($e^{+\di 2 \pi n/N}$ and $e^{-\di 2 \pi n/N}$) are applied at the left and right interfaces of the circular sectors~\cite{Thomas1979}, which leads to an eigenvalue problem $\omega(n)$: $\tilde{\bf K}(n) \tilde{\bf u}(n) = -\omega^2_n \tilde{\bf M}(n) \tilde{\bf u}(n)$, where $n$ is the order of cyclic periodicity that assumes integer and finite values such that $n \in [-N/2, ..., 0, ..., +N/2]$ for even $N$ and $n \in [-(N-1)/2, ..., 0, ..., +(N-1)/2]$ for odd $N$. Through variation of $n$, the eigenfrequencies representing the modes of a complete circular domain with free boundary conditions at the outer edges are obtained at a reduced computational cost. However, radial boundaries introduce defect modes that contaminate the band gaps. Since band gaps are characterized by low modal densities, their presence can be effectively highlighted by representing the density of states (DOS)~\cite{Duncan2020,rosa2020topological}. The DOS is given by the derivative of the mode count $M$ with respect to frequency, i.e. $\text{DOS}(\Omega)=dM/d\Omega$ ~\cite{mulhall2014calculating}, and is approximated herein by a central derivative scheme, i.e. $\text{DOS} \left( {\Omega_i} \right) = {2}/({\Omega_{i+1}-\Omega_{i-1}})$.

Figure \ref{fig:figure5} displays the spectrum for different symmetry orders as a function of volume fraction (left panels): all modal periodicities n are collapsed for each volume fraction and the frequencies are color-coded by the computed $\text{DOS}$. Dark colors are associated with dense bulk bands, while light colors are associated primarily with low-density defect modes lying inside band gaps. Bulk modes also exhibit lower density of states at low frequencies (i.e., light colors), but they are not related to the band gaps. Notably, this approach identifies possible band gaps for all symmetry orders, in particular a large low-density region is observed in all cases. A few modes identified in the spectra are displayed in Figs. a.i to d.iv. For the 6-fold case (Fig. 5a), any mode inside the gap is a defect mode localized at the edges of the circular domain (Figs. a.i to a.iii). However, for the three quasiperiodic cases presented in Fig. 5b-d, a few modes inside the assumed band gaps are localized in a region inside the wedge (not necessarily on the edges) indicating possible bulk bands. While further investigations are needed, these modes could occur due to the fractal nature of quasiperiodic media, where initially dense bulk bands may split into multiple thin bulk bands wherein localized modes in the domain may be observed~\cite{Pal2019}. Nonetheless, the gaps estimated through this approach are confirmed in frequency response computations, whereby omnidirectional harmonic forces were applied to the center of circular domains with free boundary conditions, and the response is measured through the average displacement over the external edge. The results are displayed as a function of volume fraction in the right panels of Fig.~\ref{fig:figure5}, where large attenuation regions (blue) are in agreement with the predicted gaps. Moreover, the attenuation zones of the crystals (see \ref{sec:Extra3} for the 4-fold crystal) are well defined and deeper while the attenuation zones of the quasicrystals are usually wider with smaller intensity. For each case, one representative response field is displayed for a frequency in the bulk and for a frequency inside the attenuation zones (Figs. a.I to d.II). In the bulk frequency, the displacement fields follow the corresponding rotational fold symmetry, while, in the attenuation frequency, the vibration attenuation field has no preferential direction as the symmetry order increases. This behavior is also related to the isotropy of quasiperiodic crystals that leads to nearly omnidirectional attenuation. The representative forced response functions for specific volume fractions are also provided in \ref{sec:Extra3}. The 14-fold quasicrystal possess the best vibration attenuation performance due its large single (i.e., continuum) band gap lying from low to high volume fractions. 


\section{Conclusions} \label{sec:Conclusions}
We investigated a family of quasiperiodic elastic domains with different rotational symmetry orders that are directly enforced through a design procedure in reciprocal space. Although lacking translational symmetry, the higher order rotational symmetry of quasiperiodic domains results in higher isotropic behavior when compared to periodic crystals. Indeed, results indicate that quasiperiodic composites provide an interpolation between the properties of two constituent materials with high equivalent stiffness and high isotropy for all volume fractions. Furthermore, such isotropy is also noticed in terms of their dynamic behavior, where omnidirectional wave propagation is observed at low frequencies, and retained to a certain degree also for higher frequencies. While the low-frequency wave behavior agrees with estimates obtained through equivalent homogenized properties, a rigorous characterization of the wave propagation behavior of quasiperiodic media for arbitrary frequencies is still an open question to be explored in future works. The main limitation is associated with their lack of translational symmetry which precludes Bloch's theorem to be applied. In that regard, dynamic homogenization techniques may play an important role in future studies on the dynamics of quasiperiodic domains \cite{Nassar2015, Sridhar2018, Meng2018}. In terms of band gap estimation, we investigated their spectra through an analysis based on an wedge-type unit cell combined with density of states computations, revealing band gaps whose existence were confirmed through frequency response computations. This analysis may potentially be extended to other types of quasiperiodic and generic aperiodic domains retaining rotational symmetry, where band gaps may be further explored in the context of topology and edge state phenomena. The results presented in this paper identifies a series of properties of quasiperiodic continuous elastic domains that were so far largely unexplored, opening new possibilities for the design of novel architectured materials as well as providing new opportunities for the general exploration of the physics of quasiperiodic media. Multiple opportunities are identified for future studies, such as investigating their topological properties, the role of defects and imperfections, non-linearities and, of course, experimental investigations. 


\section*{Acknowledgments}
D. Beli and C. D. Marqui gratefully acknowledge the support from S\~{a}o Paulo Research Foundation (FAPESP) through grant reference numbers: 2018/18774-6, 2019/22464-5 and 2018/15894-0 (Research project - Periodic structure design and optimization for enhanced vibroacoustic performance: ENVIBRO). M. Rosa and M. Ruzzene gratefully acknowledge the support from the National Science Foundation (NSF) through the EFRI 1741685 grant and from the Army Research Office through grant W911NF-18-1-0036.



\bibliographystyle{unsrt}
\bibliography{references}


\appendix


\section{Design strategy results for other composites} \label{sec:Extra1}

\begin{figure}[h!]
	\centering
	{\includegraphics[scale=0.600]{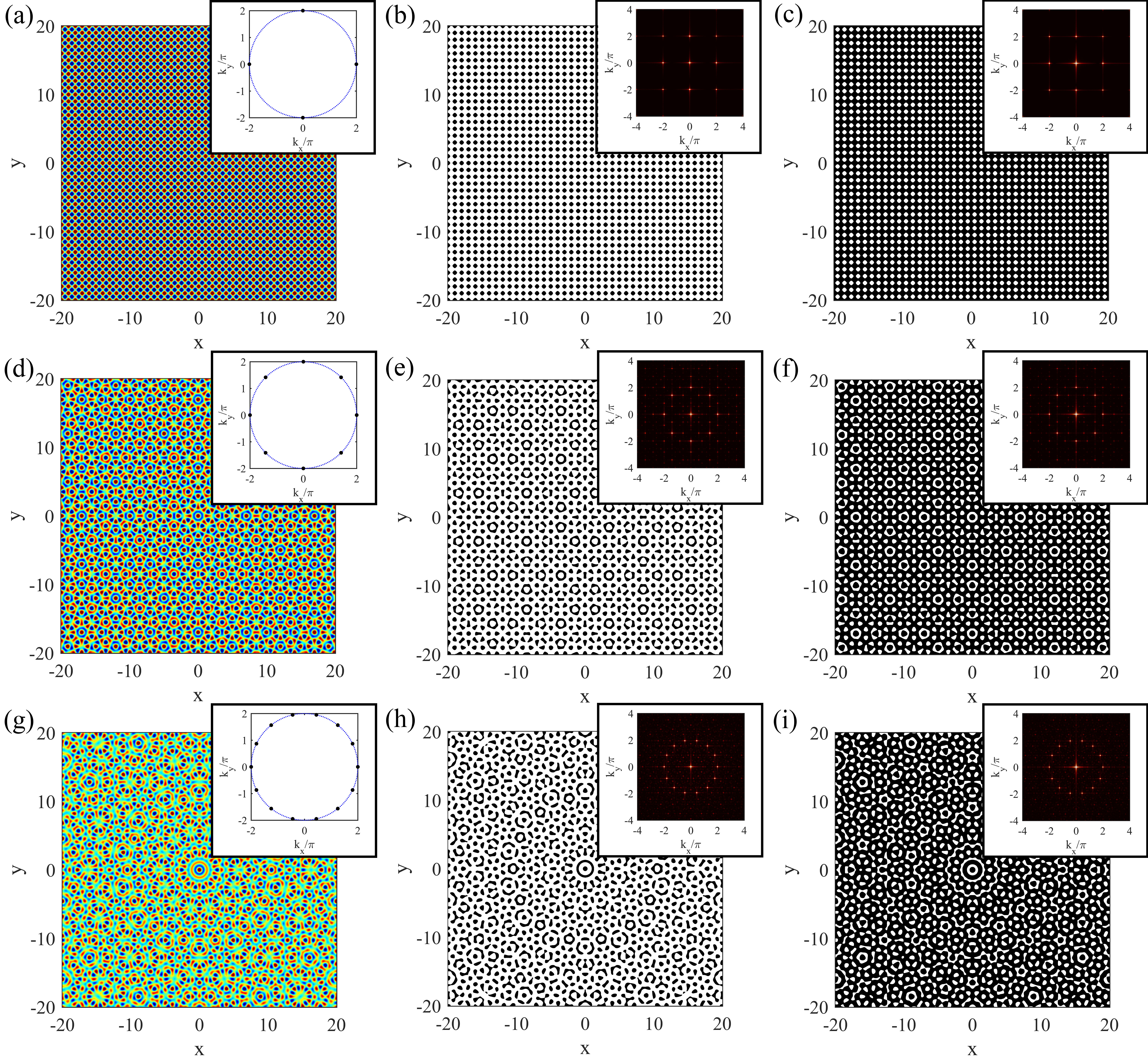}}
	\caption{\label{fig:figureSM3} Design strategy depicted in Fig. \ref{fig:figure1} exemplified for the 4-fold (a-c), 8-fold (d-f) and 14-fold (g-i): real field (a,d,g), threshold distributions with ${\tt vf} = 0.30$ (b,e,h) and threshold distributions with ${\tt vf} = 0.70$ (c,f,i).} 
\end{figure}

The design strategy (i.e, real field derivation and threshold procedure) for the other fold symmetries (4-, 8- and 14-fold) is presented in Fig. \ref{fig:figureSM3}. These composites have similar wavenumber behavior when compared to Fig. \ref{fig:figure1}, where the crystal displays a well defined Brillouin zone while the quasiperiodic crystals have a Bragg diffraction with high order symmetry according to the design fold rotation.

\section{Effective elastic properties and non-dispersive waves} \label{sec:ContMechWave}

The effective elastic tensor that relates the average stress-strain behavior is given by $\bar{\sigma}_{ij} = \bar{c}_{ijkl} \bar{\varepsilon}_{kl}$ \cite{Hollister1992}. The average strain can be written in function of the boundary displacements as
\begin{equation}
\bar{\varepsilon}_{kl} = \frac{1}{V_{\tt RVE}} \int_{V_{\tt RVE}}\varepsilon_{kl} \ dV_{\tt RVE}= \frac{1}{V_{\tt RVE}} \int_{S_{\tt RVE}} \frac{1}{2}\left(u_k n_l + u_l n_k \right) dS_{\tt RVE}, 
\label{eq:eq_sm1a}
\end{equation}
where $\bar{\varepsilon}_{kl}$ is the average strain, ${\varepsilon}_{kl}$ is the local strain, $u_k$ is the displacement on the $\tt RVE$ boundary, $n_k$ is the normal vector to the $\tt RVE$ boundary, and $S_{\tt RVE}$ is the $\tt RVE$ boundary. Moreover, the average stress can be written in function of the boundary tractions as
\begin{equation}
\bar{\sigma}_{ij} = \frac{1}{V_{\tt RVE}} \int_{V_{\tt RVE}} \sigma_{ij} \ dV_{\tt RVE}= \frac{1}{V_{\tt RVE}} \int_{S_{\tt RVE}} \frac{1}{2}\left(t_i x_j + t_j x_i \right) dS_{\tt RVE},
\label{eq:eq_sm1b}
\end{equation}
where $\bar{\sigma}_{ij}$ is the average stress, $\sigma_{ij}$ is the local stress, $t_i$ is the traction on the $\tt RVE$ boundary and $x_i$ is the local coordinates of the RVE boundary. For a linear elastic orthotropic model in plane strain behavior, only some components of the elastic tensor are required to describe the stress-strain relation, which is reduced to  
\begin{equation}
\begin{bmatrix} \bar{\sigma}_{xx} \\ \bar{\sigma}_{yy} \\ \bar{\sigma}_{xy}\end{bmatrix} =  \begin{bmatrix} \bar{c}_{11} & \bar{c}_{12} & 0 \\ \bar{c}_{21} & \bar{c}_{22} & 0 \\ 0 & 0 & \bar{c}_{66} \end{bmatrix} \begin{bmatrix} \bar{\varepsilon}_{xx} \\ \bar{\varepsilon}_{yy} \\ \bar{\varepsilon}_{xy}\end{bmatrix}.
\label{eq:eq_sm1}
\end{equation}

\begin{figure}[h!]
	\centering
    {\includegraphics[scale=0.450]{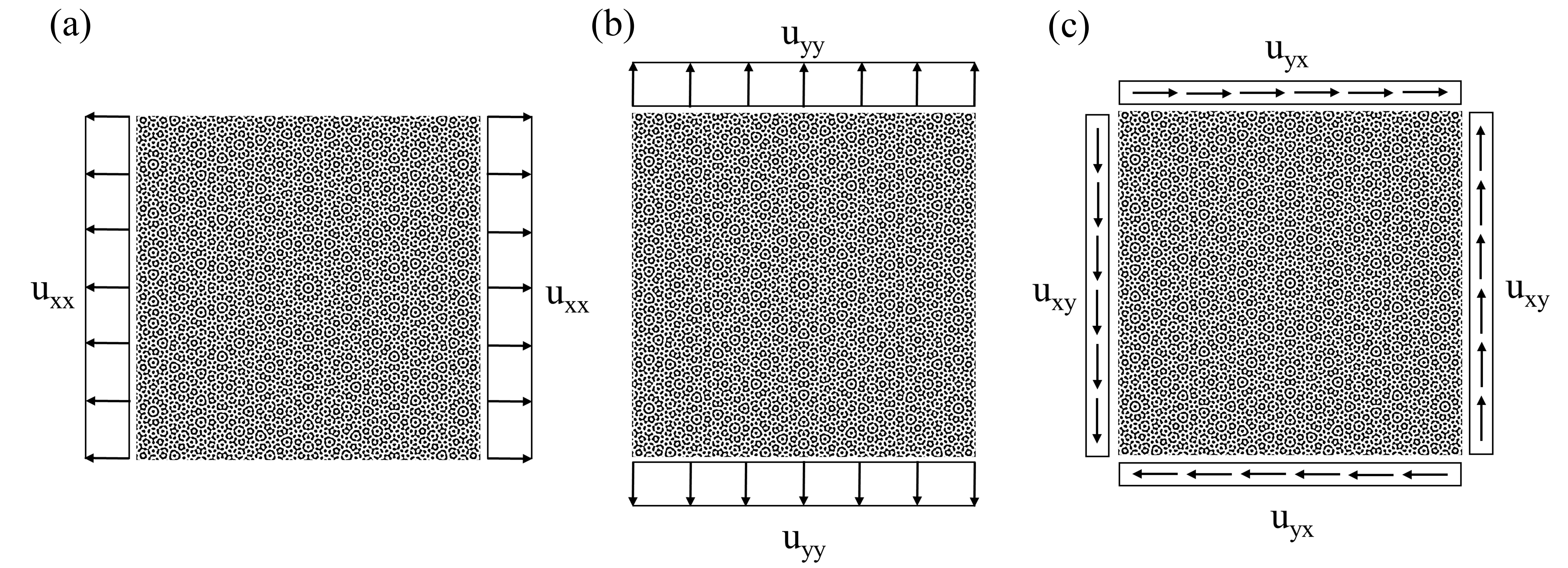}} 
	\caption{\label{fig:figureSM1} Standard mechanics approach \cite{Hollister1992}: axial load at $x$ direction (a), axial load at $y$ direction (b) and shear load (c).} 
\end{figure}

By applying three independent displacement loads over the $\tt RVE$ as shown in Fig. \ref{fig:figureSM1}, the resulting average strain and stress can be computed , and hence, the components of the elastic tensor are derived \cite{Hollister1992, Nguyen2012}. From the independent axial loads aligned to $x$ axis and to $y$ axis, respectively,
\begin{equation}
\bar{\nu}_{xy} = - \frac{\bar{\varepsilon}_{yy}^{\ x}}{\bar{\varepsilon}_{xx}^{\ x}} =  \frac{c_{21}}{c_{22}} \quad \text{and} \quad \bar{\nu}_{yx} =  - \frac{\bar{\varepsilon}_{xx}^{\ y}}{\bar{\varepsilon}_{yy}^{\ y}} =  \frac{c_{12}}{c_{11}},
\label{eq:eq_sm2a}
\end{equation}
where $\nu_{ij}$ correspond to traction in direction $j$ when the extension is applied in direction $i$. Therefore, from the computed average stress, Eq. \ref{eq:eq_sm1} and Eq. \ref{eq:eq_sm2a},
\begin{subequations}
\begin{align}
\bar{\sigma}_{xx}^{\ x} = \bar{c}_{11}\bar{\varepsilon}_{xx}^{\ x}+\bar{c}_{12}\bar{\varepsilon}_{yy}^{\ x} = \bar{c}_{11} \left(1- \bar{c}_{12} \bar{c}_{21} / \left( \bar{c}_{22} \bar{c}_{11} \right) \right) \bar{\varepsilon}_{xx}^{\ x} \rightarrow \bar{c}_{11} = \left( \bar{\sigma}_{xx}^{\ x} / \bar{\varepsilon}_{xx}^{\ x} \right)/ \left( 1 - \bar{\nu}_{xy} \bar{\nu}_{yx}\right), \\
\bar{\sigma}_{yy}^{\ y} = \bar{c}_{22}\bar{\varepsilon}_{yy}^{\ y}+\bar{c}_{21}\bar{\varepsilon}_{xx}^{\ y} = \bar{c}_{22} \left(1- \bar{c}_{12} \bar{c}_{21} / \left( \bar{c}_{22} \bar{c}_{11} \right) \right) \bar{\varepsilon}_{yy}^{\ y} \rightarrow \bar{c}_{22} = \left( \bar{\sigma}_{yy}^{\ y} / \bar{\varepsilon}_{yy}^{\ y} \right)/ \left( 1 - \bar{\nu}_{xy} \bar{\nu}_{yx}\right). 
\end{align} 
\label{eq:eq_sm2b}  
\end{subequations}
And from the shear load $\bar{c}_{66} = \bar{\sigma}_{xy}/ \bar{\varepsilon}_{xy}$
Through the tensor transformation by means of direction cosines, the directional dependence of the effective elastic coefficients can be computed  $\bar{c}'_{ijkl}(\theta) = \bar{c}_{mnpq} a_{im}a_{jn}a_{kp}a_{lq}$ \cite{Rosario2017, Portela2020}. And hence, the directional dependence of the elastic properties 
\begin{equation}
\bar{E} (\theta) =  \bar{c}_{11} (\theta) \left(1-\frac{\bar{c}_{12} (\theta) \bar{c}_{21} (\theta)}{\bar{c}_{22} (\theta) \bar{c}_{11} (\theta)} \right), \quad \bar{G} (\theta) = \bar{c}_{66} (\theta) \quad \text{and} \quad \bar{\nu}(\theta) = \frac{\bar{c}_{21}(\theta)}{\bar{c}_{22}(\theta)}
\label{eq:eq_sm3}
\end{equation}

Finally, the components of the elastic tensor can also be related to the non-dispersive (i.e., low frequency assumption) longitudinal and shear waves by the group velocity or phase velocity 
\begin{equation}
\bar{c}_{L}(\theta) = \sqrt{\frac{\bar{c}_{11}(\theta)}{\bar{\rho}}}, \quad \text{and} \quad
\bar{c}_{S}(\theta) = \sqrt{\frac{\bar{c}_{66}(\theta)}{\bar{\rho}}},
\label{eq:eq_sm4}
\end{equation}  
where $\bar{c}_L$ is the longitudinal speed of the P-wave, $\bar{c}_S$ is the shear speed of the S-wave, and the equivalent mass density is given by $ \bar{\rho} = \rho_{\tt 0} \left( 1-{\tt vf} \right) +\rho_{\tt 1} {\tt vf} $. 

\section{Dispersion analysis for the 4-fold composite} \label{sec:Extra2}

\begin{figure}[h!]
	\centering
    {\includegraphics[scale=0.65]{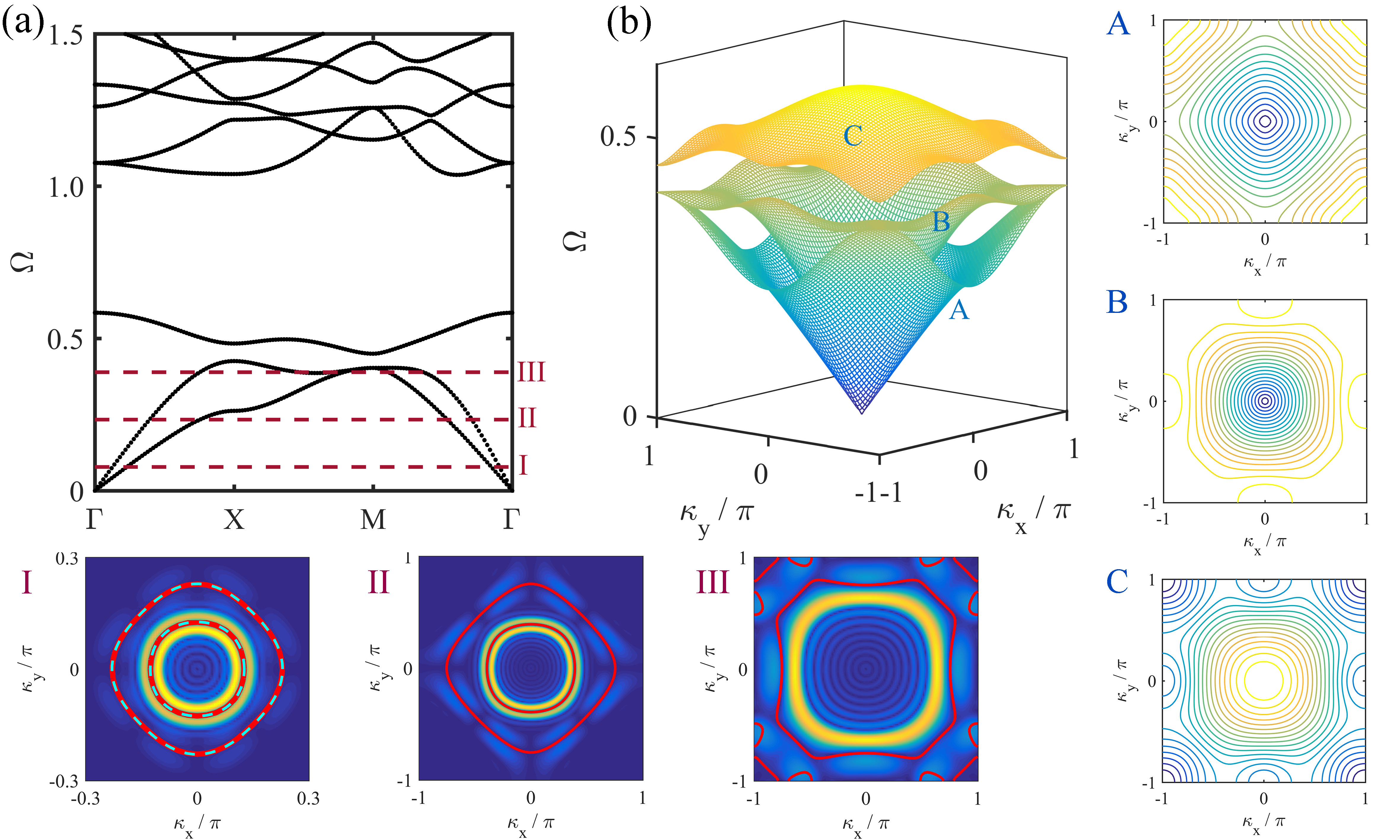}} 
	\caption{\label{fig:figureSM4fold} Dispersion analysis for the 4-fold composite. Band structure (a), where the excitation frequencies for transient analysis are depicted by straight red lines from I to III; moreover, for these frequencies are shown in (I-III): the iso-frequency contours from the dispersion surfaces (red lines), Fourier transform from transient simulation $|\hat{\bf U}(\kappa_x,\kappa_y,\omega)|$ computed in Section \ref{sec:Dynamical} (color map) as well as the P-wave and S-wave contours obtained from the effective properties computed in Section \ref{sec:Mechanical} (dashed blued lines). Dispersion surfaces (b) with the iso-frequency contours for the first (A), second (B) and third (C) bands.}  
\end{figure}

In contrast with the quasiperiodic composites, the periodic composites such as the 4-fold and the 6-fold present a well defined Brillouin zone. Therefore, Bloch periodic conditions can be enforced on their unit cell boundaries to investigate their dispersion features such as band gaps and directionality. We here present the dispersion surfaces and contours for the 4-fold configuration with ${\tt vf} =0.30$ (Fig. \ref{fig:figureSM4fold}). These results are not presented in the main text since they cannot be extended to the quasicrystals, however they are used to guide the transient analysis for the quasiperiodic symmetries with same volume fraction. As highlighted in Fig.~\ref{fig:figureSM4fold}(a), the three frequencies considered for the transient analysis (dashed red lines) are chosen to sample the first two bands, which are known to present dominant longitudinal and shear wave behavior. The dispersion surfaces and their contours for the first three bands are also presented, highlighting the anisotropic behavior expected from the 4-fold configuration. Finally, the contours for the selected frequencies I-III are overlaid (in red lines) to the 2D ffts obtained for the transient computations, demonstrating very good agreement. Note also that for low frequencies (case I), the dispersion contours obtained through Bloch analysis are in very good agreement with the contours estimated through the effective mechanical properties (dashed blue lines), which further validates that approach.

\section{Harmonic response at specific volume fractions} \label{sec:Extra3}

\begin{figure}[h!]
	\centering
	{\includegraphics[width=1\textwidth]{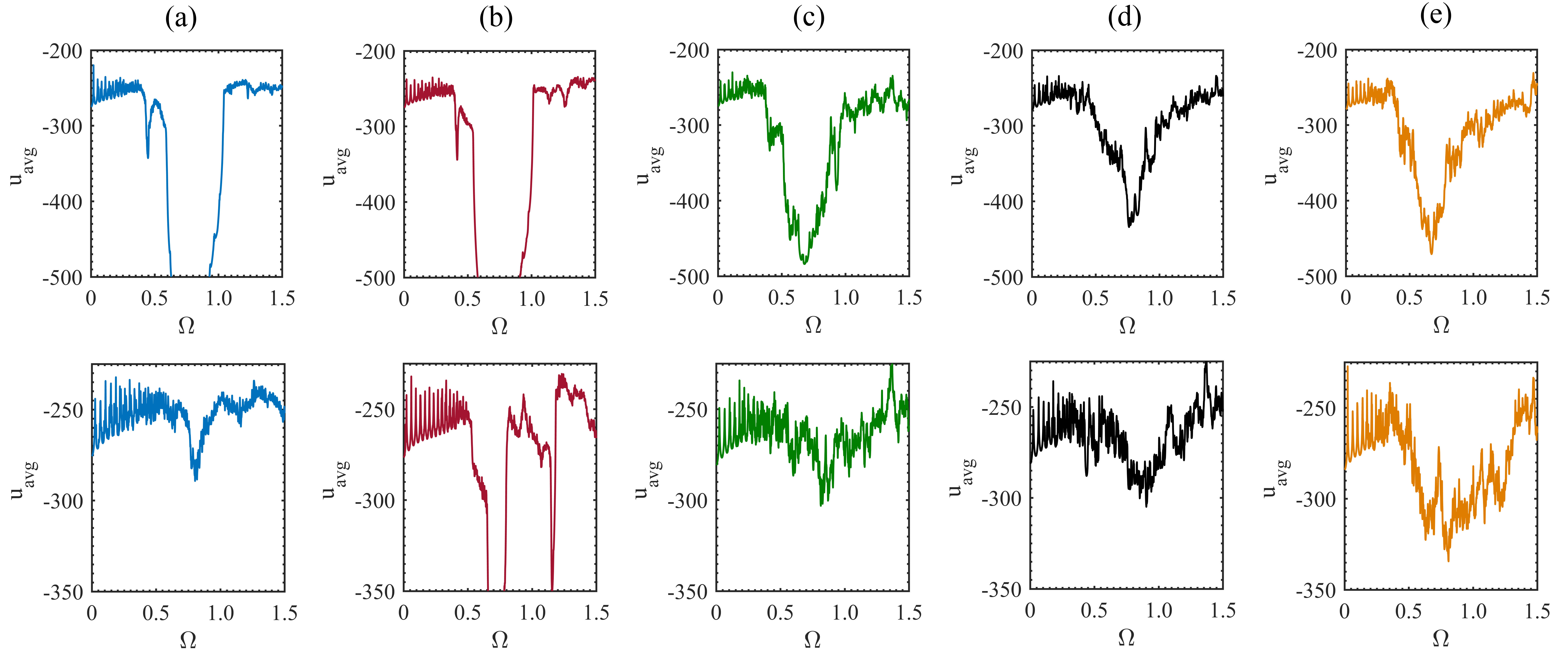}}
	\caption{\label{fig:figureSM6} Average harmonic displacement response over the external edge with ${\tt vf} = 0.30$ (top line) and ${\tt vf} = 0.70$ (bottom line): 4- (a), 6- (b), 8- (c), 10- (d) and 14- (e) fold rotations.} 
\end{figure}

In Fig. \ref{fig:figureSM6} complements the results in Fig. \ref{fig:figure5} by showing the forced response for the different fold orders in specific volume fractions: the crystals present a well defined vibration attenuation profile with high intensity, while the quasicrystals have a jagged vibration attenuation profile with low intensity.

\end{document}